\documentclass[article]{elsarticle}

\usepackage{lineno,hyperref}
\usepackage{algorithm}
\usepackage{algpseudocode}
\usepackage{pdflscape}
\usepackage{booktabs}
\usepackage{natbib}
\usepackage{subfig}

\journal{Journal of \LaTeX\ Templates}

\algblockdefx{MAP}{ENDMAP}[1]%
{\textbf{map} #1}%
{\textbf{end map}}

\algnewcommand{\IIf}[1]{\State\algorithmicif\ #1\ \algorithmicthen}
\algnewcommand{\EndIIf}{\unskip\ \algorithmicend\ \algorithmicif}

\begin{document}

\begin{frontmatter}

\title{Enabling Smart Data: Noise filtering in Big Data classification}

\author[granada]{Diego Garc\'ia-Gil\corref{mycorrespondingauthor}}
\ead{djgarcia@decsai.ugr.es}

\author[granada]{Juli\'an Luengo}
\ead{julianlm@decsai.ugr.es}

\author[granada]{Salvador Garc\'ia}
\ead{salvagl@decsai.ugr.es}

\author[granada]{Francisco Herrera}
\ead{herrera@decsai.ugr.es}

\address[granada]{Department of Computer Science and Artificial Intelligence, University of Granada, CITIC-UGR, Granada, Spain, 18071}
\cortext[mycorrespondingauthor]{Corresponding author}

\begin{abstract}
In any knowledge discovery process  the value of extracted knowledge is directly related to the quality of the data used.
Big Data problems, generated by massive growth in the scale of data observed in recent years, also follow the same dictate.
A common problem affecting data quality is the presence of noise, particularly in classification problems, where label noise refers to the incorrect labeling of training instances, and is known to be a very disruptive feature of data.
However, in this Big Data era, the massive growth in the scale of the data poses a challenge to traditional proposals created to tackle noise, as they have difficulties coping with such a large amount of data.
New algorithms need to be proposed to treat the noise in Big Data problems, providing high quality and clean data, also known as Smart Data.
In this paper, two Big Data preprocessing approaches to remove noisy examples are proposed: an homogeneous ensemble and an heterogeneous ensemble filter, with special emphasis in their scalability and performance traits.
The obtained results show that these proposals enable the practitioner to efficiently obtain a Smart Dataset from any Big Data classification problem.
\end{abstract}


\end{frontmatter}


\section{Introduction}

Vast amounts of information surround us today.
Technologies such as the Internet generate data at an exponential rate thanks to the affordability and great development of storage and network resources. 
It is predicted that by 2020, the digital universe will be 10 times as big as it was in 2013, totaling an astonishing 44 zettabytes~\cite{IDC}.
The current volume of data has exceeded the processing capabilities of classical data mining systems~\cite{Wu2014} and have created a need for new frameworks for storing and processing this data.
It is widely accepted that we have entered the Big Data era~\cite{mayer-schonberger13revolution}.
Big Data is the set of technologies that make processing such large amounts of data possible~\cite{PHILIPCHEN}, while most of the classic knowledge extraction methods cannot work in a Big Data environment because they were not conceived for it.

Big Data as concept is defined around five aspects: data volume, data velocity, data variety, data veracity and data value~\cite{laney01controlling3v}. 
While the volume, variety and velocity aspects refer to the data generation process and how to capture and store the data, veracity and value aspects deal with the quality and the usefulness of the data.
These two last aspects become crucial in any Big Data process, where the extraction of useful and valuable knowledge is strongly influenced by the quality of the used data. 

In Big Data, the usage of traditional preprocessing techniques~\cite{garcia2015,1999-Pyle,García2016} to enhance the data is even more time consuming and resource demanding, being unfeasible in most cases.
The lack of efficient and affordable preprocessing techniques implies that the problems in the data will affect the models extracted.
Among all the problems that may appear in the data, the presence of \emph{noise} in the dataset is one of the most frequent.
Noise can be defined as the partial or complete alteration of the information gathered for a data item, caused by an exogenous factor not related to the distribution that generates the data.
Learning from noisy data is an important topic in machine learning, data mining and pattern recognition, as real world data sets may suffer from imperfections in data acquisition, transmission, storage, integration and categorization.
Noise will lead to excessively complex models with deteriorated performance~\cite{Wu08Miningwith}, resulting in even larger computing times for less value.

The impact of noise in Big Data, among other pernicious traits, has not been disregarded. 
Recently, Smart Data (focusing on veracity and value) has been introduced, aiming to filter out the noise and to highlight the valuable data, which can be effectively used by companies and governments for planning, operation, monitoring, control, and intelligent decision making.
Three key attributes are needed for data to be smart, it must be accurate, actionable and agile:
\begin{itemize}
	\item Accurate: data must be what it says it is with enough precision to drive value. Data quality matters.
	\item Actionable: data must drive an immediate scalable action in a way that maximizes a business objective like media reach across platforms. Scalable action matters.
	\item Agile: data must be available in real-time and ready to adapt to the changing business environment. Flexibility matters.
\end{itemize}

Advanced Big Data modeling and analytics are indispensable for discovering the underlying structure from retrieved data in order to acquire Smart Data.
In this paper we provide several preprocessing techniques for Big Data, transforming raw, corrupted datasets into Smart Data.
We focus our interest on classification tasks, where two types of noise are distinguished: \emph{class noise}, when it affects the class label of the instances, and \emph{attribute noise}, when it affects the rest of attributes. The former is known to be the most disruptive~\cite{2014-KAIS-Saez,Zhu04classnoisevsattributenoise}. 
Consequently, many recent works, including this contribution, have been devoted to resolving this problem or at least to minimize its effects (see~\cite{2014-TNNLS-Frenay} for a comprehensive and updated survey).
While some architectural designs are already proposed in the literature\cite{zerhari2016class}, there is no particular algorithm which deals with noise in Big Data classification, nor a comparison of its effect on model generalization abilities or computing times.

Thereby we propose a framework for Big Data under Apache Spark for removing noisy examples composed of two algorithms based on ensembles of classifiers.
The first one is an homogeneous ensemble, named Homogeneous Ensembe for Big Data (HME-BD), which uses a single base classifier (Random Forest~\cite{Breiman2001}) over a partitioning of the training set.
The second ensemble is an heterogeneous ensemble, namely Heterogeneous Ensembe for BigData (HTE-BD), that uses different classifiers to identify noisy instances: Random Forest, Logistic Regression and K-Nearest Neighbors (KNN) as base classifiers.
For the sake of a more complete comparison, we have also considered a simple filtering approach based on similarities between instances, named Edited Nearest Neighbor for Big Data (ENN-BD).
ENN-BD examines the nearest neighbors of every example in the training set and eliminates those whose majority of neighbors belong to a different class.
All these techniques have been implemented under the Apache Spark framework~\cite{spark,spark2} and can be downloaded from the Spark's community repository~\footnote{\url{https://spark-packages.org/package/djgarcia/NoiseFramework}}.

To show the performance of the three proposed algorithms, we have carried out an experimental evaluation with four large datasets, namely \textit{SUSY}, \textit{HIGGS}, \textit{Epsilon} and \textit{ECBDL14}.
We have induced several levels of class noise to evaluate the effects of applying such framework and the improvements obtained in terms of classification accuracy for two classifiers: a decision tree and the KNN technique.
Decision trees with pruning are known to be tolerant to noise, while KNN is a noise sensitive algorithm when the number of selected neighbors is low.
These differences allow us to better compare the effect of the framework in classifiers which behave differently towards noise.
We also show that, for the Big Data problems considered, the classifiers also benefit from applying the noise treatment even when no additional noise is induced, since Big Data problems contain implicit noise due to incidental homogeneity, spurious correlations and the accumulation of noisy examples~\cite{Fan2014}.
The results obtained indicate that the framework proposed can successfully deal with noise.
In particular, the homogeneous ensemble is a suitable technique for dealing with noise in Big Data problems, with low computing times and enabling the classifier to achieve better accuracy.

The remainder of this paper is organized as follows: Section~\ref{sec:background} presents the concepts of noise, MapReduce and Smart Data. Section~\ref{sec:proposal} explains the proposed framework. Section~\ref{sec:experiments} describes the experiments carried out to check the performance of the framework. Finally, Section~\ref{sec:conclusions} concludes the paper.

\section{Related work}
\label{sec:background}

In this section we first present the problem of noise in classification tasks in Section~\ref{sec:noise}.
Then we introduce the MapReduce framework commonly used in Big Data solutions in Section~\ref{sec:mapreduce}.
Finally, we provide an insight into Smart Data in~\ref{sec:smartdata}.

\subsection{Class noise vs. attribute noise}
\label{sec:noise}

In a classification problem, several effects of this noise can be observed by analyzing its spatial characteristics: noise may create small clusters of instances of a particular class in the instance space corresponding to another class, displace or remove instances located in key areas within a concrete class, or disrupt the boundaries of the classes resulting in an increased boundaries overlap.
All these imperfections may harm data interpretation, the design, size, building time, interpretability and accuracy of models, as well as decision making~\cite{Zhong04AnalyzingSoftware, Zhu04classnoisevsattributenoise}.

As described by {Wang et al.}~\cite{Wu96KnowledgeAcquisition}, from the large number of components that comprise a dataset, class labels and attribute values are two essential elements in classification datasets.
Thus, two types of noise are commonly differentiated in the literature~\cite{Zhu04classnoisevsattributenoise,Wu96KnowledgeAcquisition}:
\begin{itemize}
	
	\item \emph{Class noise}, also known as \emph{label noise}, takes place when an example is wrongly labeled.
	Class noise includes contradictory examples~\cite{Teng99CorrectingNoisy,2014-KAIS-Saez} (examples with identical input attribute values having different class labels) and misclassifications~\cite{Zhu04classnoisevsattributenoise} (examples which are incorrectly labeled).
	
	\item \emph{Attribute noise} refers to corruptions in the values of the input attributes.
	It includes erroneous attribute values, missing values and incomplete attributes or ``do not care'' values.
	Missing values are usually considered independently in the literature, so \emph{attribute noise} is mainly used for erroneous values~\cite{Zhu04classnoisevsattributenoise}.
	
\end{itemize}

Class noise is generally considered more harmful to the learning process, and methods for dealing with class noise are more frequent in the literature~\cite{Zhu04classnoisevsattributenoise}.
Class noise may have many reasons, such as errors or subjectivity in the data labeling process, as well as the use of inadequate information for labeling. 
Data labeling by domain experts is generally costly, and automatic taggers are used (e.g., sentiment analysis polarization~\cite{liu2015sentiment}), increasing the probability of class noise. 

Due to the increasing attention from researchers and practitioners, numerous techniques have been developed to tackle it~\cite{2014-TNNLS-Frenay,Zhu04classnoisevsattributenoise,garcia2015}. 
These techniques include learning algorithms robust to noise as well as data preprocessing techniques that remove or ``repair'' noisy instances.
In~\cite{2014-TNNLS-Frenay} the mechanisms that generate label noise are examined, relating them to the appropriate treatment procedures that can be safely applied:
\begin{itemize}
	
	\item On the one hand, \emph{algorithm level} approaches attempt to create robust classification algorithms that are little influenced by the presence of noise. 
	This includes approaches where existing algorithms are modified to cope with label noise by either being modeled in the classifier construction~\cite{lawrence2001,Li2007}, by applying pruning strategies to avoid overfiting as in~\cite{Quinlan93C45Programs} or by diminishing the importance of noisy instances with respect to clean ones~\cite{Miao2016}. 
	Recent proposals exist which that combine these two approaches, which model the noise and give less relevance to potentially noisy instances in the classifier building process~\cite{Bouveyron2009}.
	
	\item On the other hand, \emph{data level} approaches (also called \emph{filters}) try to develop strategies to cleanse the dataset as a previous step to the fit of the classifier, by either creating ensembles of classifiers~\cite{Brodley99identifyingmislabeled}, iteratively filtering noisy instances~\cite{Khoshgoftaar07ImprovingSoftware}, computing metrics on the data or even hybrid approaches that combine several of these strategies.
	
\end{itemize}

In the Big Data environment there is a special need for noise filter methods. 
It is well known that the high dimensionality and example size generate accumulated noise in Big Data problems~\cite{Fan2014}.
Noise filters reduce the size of the datasets and improve the quality of the data by removing noisy instances, but most of the classic algorithms for noisy data, noise filters in particular, are not prepared for working with huge volumes of data.

\subsection{Big Data. MapReduce and Apache Spark} 
\label{sec:mapreduce}

The globalization of the Big Data paradigm is generating a large response in terms of technologies that must deal with the rapidly growing rates of generated data~\cite{WANG2016747}.
Among all of them, MapReduce is the seminal framework designed by Google in 2003 ~\cite{Dean:2008:MSD:1327452.1327492}. It follows a divide and conquer approach to process and generate large datasets with parallel and distributed algorithms on a cluster. The MapReduce model is composed of two phases: Map and Reduce.
The Map phase performs a transformation of the data, and the Reduce phase performs a summary operation. Briefly explained, first the master node splits the input data and distributes it across the cluster. Then the Map transformation is applied to each key-value pair in the local data. Once that process is finished the data is redistributed based on the key-value pairs generated in the Map phase. Once all pairs belonging to one key are in the same node, it is processed in parallel. Apache Hadoop~\cite{White:2012:HDG:2285539}~\cite{hadoop} is the most popular open-source framework based on the MapReduce model.

Apache Spark~\cite{spark,spark2} is an open-source framework for Big Data processing built around speed, ease of use and sophisticated analytics. Its main feature is its ability to use in-memory primitives. Users can load their data into memory and iterate over it repeatedly, making it a suitable tool for ML algorithms. The motivation for developing Spark came from the limitations in the MapReduce/Hadoop model~\cite{DBLP:journals/corr/abs-1209-2191, WIDM:WIDM1134}:

\begin{itemize}
	\item Intensive disk usage
	\item Insufficiency for in-memory computation
	\item Poor performance on online and iterative computing.
	\item Low inter-communication capacity.
\end{itemize}

Spark is built on top of a distributed data structure called Resilient Distributed Datasets (RDDs)~\cite{Zaharia:2012:RDD:2228298.2228301}. Operations on RDDs are applied to each partition of the node local data. RDDs support two types of operations: transformations, which are not evaluated when defined and produce a new RDD, and actions, which evaluate all the previous transformations and return a new value. The RDD structure allows programmers to persist them into memory or disk for re-usability purposes. RDDs are immutable and fault-tolerant by nature. All operations are tracked using a "lineage", so that each partition can be recalculated in case of failure.

Although new promising frameworks for Big Data are emerging, like Apache Flink~\cite{flink}, Apache Spark is becoming the reference in performance~\cite{García-Gil2017}.

\subsection{From Big Data to Smart Data}
\label{sec:smartdata}

Big Data is an appealing discipline that presents an immense potential for global economic growth and promises to enhance competitiveness of high technological countries.
Such as occurs in any knowledge extraction process, vast amounts of data are analyzed, processed, and interpreted in order to generate profits in terms of either economic or advantages for society.
Once the Big Data has been analyzed, processed, interpreted and cleaned, it is possible to access it in a structured way. 
This transformation is the difference between ``Big'' and ``Smart'' Data~\cite{Lenk2015}.

The first step in this transformation is to perform an integration process, where the semantics and domains from several large sources are unified under a common structure. 
The usage of ontologies to support the integration is a recent approach~\cite{Fadili2016,Chen2017}, but graph databases are also an option where the data is stored in a relational form, as in healthcare domains~\cite{Raja2015}.
Even when the integration phase ends, the data is still far from being ``smart'': the accumulated noise in Big Data problems creates problems in classical Data Mining techniques, specially when the dimensionality is large~\cite{fan2008high}.
Thus, in order to be ``smart'', the data still needs to be cleaned even after its integration, and data preprocessing is the set of techniques utilized to encompass this task~\cite{garcia2015,GarciaLH16}.

Once the data is ``smart'', it can hold the valuable data and allows interactions in ``real time'', like transactional activities and other Business Intelligence applications.
The goal is to evolve from a data-centered organization to a learning organization, where the focus is set on the knowledge extracted instead of struggling with the data management~\cite{Iafrate2014}.
However, Big Data generates great challenges to achieve this since its high dimensionality and large example size imply noise accumulation, algorithmic instability and the massive sample pool is often aggregated from heterogeneous sources~\cite{Fan2014}.
While feature selection, discretization or imbalanced algorithms to cope with the high dimensionality have drawn the attention of current Big Data frameworks (such as Spark's MLlib~\cite{spark2016}) and researchers~\cite{Ramirez-Gallego17,tan14a,ramirez2016data,triguero2015rosefw}, algorithms to clean noise are still a challenge.
In summary, challenges are still present to fully operate a transition between Big Data to Smart Data.
In this paper we provide an automated preprocessing framework to deal with class noise, enabling the practitioner to reach Smart Data.

\section{Towards Smart Data: Noise filtering for Big Data}
\label{sec:proposal}

{In this section, we present the framework for Big Data under Apache Spark for removing noisy examples based on the MapReduce paradigm, proving its performance over real-world large problems.
It is a MapReduce design where all the noise filter processes are performed in a distributed way.

{In Section~\ref{sec:primitives} we describe the Spark primitives used for the implementation of the framework.
We have designed two algorithms based on ensembles. Both perform a $k$-fold on the training data, learn a model on the training partition and clean noisy instances in the test partition.
The first one is an homogeneous ensemble using Random Forest as a classifier, named HME-BD (Section~\ref{sec:hme_bd}). 
The second one, named HTE-BD (Section~\ref{sec:hte_bd}) is a heterogeneous ensemble based on the use of three different classifiers: Random Forest, Logistic Regression and KNN.
We have also implemented a simple filter based on the similarity between the instances, named ENN-BD (Section~\ref{sec:enn_bd}).

\subsection{Spark Primitives}
\label{sec:primitives}

For the implementation of the framework, we have used some basic Spark primitives from Spark API. These primitives offer much complex operations by extending the MapReduce paradigm. Here, we outline those more relevant to the algorithms \footnote{For a complete description of Spark's operations, please refer to Spark's API: \url{http://spark.apache.org/docs/latest/api/scala/index.html}}:

\begin{itemize}
	\item $map$: Applies a transformation to each element of a RDD. Once the operation has been performed to each element, the resulting RDD is returned.
	\item $zipWithIndex$: for each element of a RDD, a pair consisting in the element and its index is created, starting at 0. The resulting RDD is then returned.
	\item $join$: Return a RDD containing all pairs of elements with matching keys between two RDDs.
	\item $filter$: Return a new RDD containing only the elements that satisfy a predicate.
	\item $union$: Return a RDD of pairs as result of the union of two RDDs.
	\item $kFold$: Returns a list of $k$ pairs of RDDs with the first element of each pair containing the $train$ data, a complement of the $test$ data, and the second element containing the $test$ data, being a unique 1/kth of the data. Where $k$ is the number of folds.
	\item $randomForest$: Method to learn a Random Forest model for classification problems.
	\item $predict$: Returns a RDD containing the features and the predicted labels for a given dataset using the learned model.
	\item $learnClassifiers$: Although its not a pure Spark primitive, we use it to simplify the description of the algorithms. This primitive learns a Random Forest, Logistic Regression and 1NN models from the input data.
\end{itemize}

These Spark primitives from Spark API are used in the following sections where HME-BD, HTE-BD and ENN-BD algorithms are described.

\subsection{Homogeneous Ensemble: HME-BD}
\label{sec:hme_bd}

\begin{algorithm}[t]
	\floatname{algorithm}{Algorithm}
	\caption{HME-BD Algorithm}
	\label{cvcf}
	\begin{algorithmic}[1]
		\State \textbf{Input:} \textit{data} a RDD of tuples (label, features)
		\State \textbf{Input:} \textit{P} the number of partitions
		\State \textbf{Input:} \textit{nTrees} the number of trees for Random Forest
		\State \textbf{Output:} the filtered RDD without noise
		\State $partitions \gets kFold(data, P)$
		\State $filteredData \gets \emptyset$
		\ForAll{$train, test$ in $partitions$}
		\State $rfModel \gets randomForest(train, nTrees)$
		\State $rfPred \gets predict(rfModel, test)$
		\State $joinedData \gets join(zipWithIndex(test), zipWithIndex(rfPred))$
		\State $markedData \gets$ \MAP $original, prediction \in joinedData$
		\If {$label(original) = label(prediction)$}
		\State $original$
		\Else
		\State $(label = \emptyset, features(original))$
		\EndIf
		\ENDMAP
		\State $filteredData \gets union(filteredData, markedData)$
		\EndFor
		\State $return (filter(filteredData, label \neq \emptyset))$
	\end{algorithmic}
\end{algorithm}

The homogeneous ensemble is inspired by Cross-Validated Committees Filter (CVCF)~\cite{VA2003}. This filter removes noisy examples by partitioning the data in $P$ subsets of equal size. Then, a decision tree, such as C4.5, is learned $P$ times, each time leaving out one of the subsets of the training data. This results in $P$ classifiers which are used to predict all the training data $P$ times. Then, using a voting strategy, misclassified instances are removed.

HME-BD is also based on a partitioning scheme of the training data. There is an important difference with respect to CVCF: the use of Spark's implementation of Random Forest instead a of a decision tree as a classifier. CVCF creates an ensemble from partitioning of the training data. HME-BD also partitions the training data, but the use of Random Forest allows us to improve the voting step:

\begin{itemize}
	\item CVCF predicts the whole dataset $P$ times. We only predict the instances of the partition that Random Forest has not seen while learning the model. This step is repeated $P$ times. With this change we not only improve the performance, but also the computing time of the algorithm since it only has to predict a small part of the training data each iteration.
	\item We don't need to implement a voting strategy, the decision of whether an instance is noisy is associated with the Random Forest prediction.
\end{itemize}


Algorithm \ref{cvcf} describes the noise filtering process in HME-BD:
\begin{itemize}
	\item The algorithm filters the noise in a dataset by performing a $kFold$ on the training data. As stated previously, Spark's $kFold$ function returns a list of $(train, test)$ for a given $P$, where $test$ is a unique 1/kth of the data, and $train$ is a complement of the $test$ data.
	\item We iterate through each partition, learning a Random Forest model using the $train$ as input data and predicting the $test$ using the learned model.
	\item In order to join the $test$ data and the predicted data for comparing the classes, we use the $zipWithIndex$ operation in both RDDs. With this operation, we add an index to each element of both RDDs. This index is used as key for the join operation.
	\item The next step is to apply a Map function to the previous RDD in order to check for each instance the original class and the predicted one. If the predicted class and the original are different, the instance is marked as noise.
	\item The result of the previous Map function is a RDD where noisy instances are marked. These instances are finally removed using a $filter$ function and the resulting dataset is returned.
\end{itemize}
The following are required as input parameters: the dataset ($data$), the number of partitions ($P$) and the number of trees for the Random Forest ($nTrees$).

In Figure~\ref{fig:hme} we can see a flowchart of the HME-BD noise filtering process.

\begin{figure}
	\centering
	\hspace*{-1.5in}
	\includegraphics[scale=0.13]{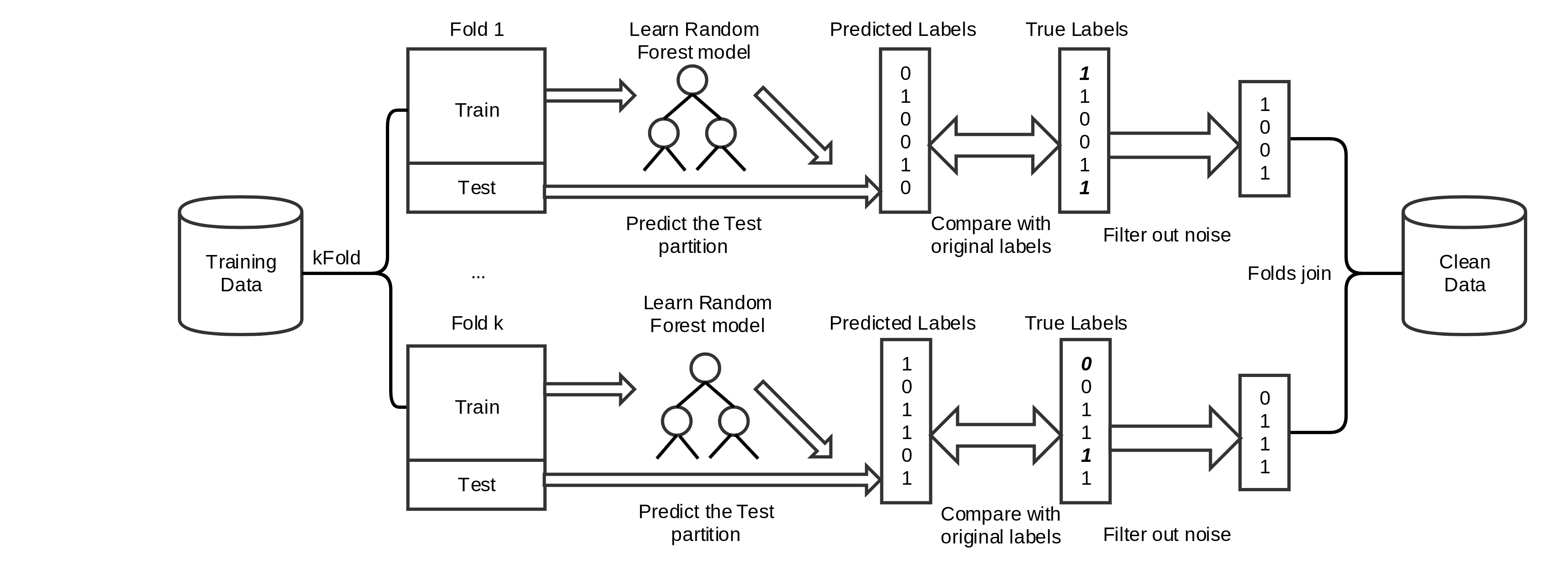}
	\caption{HME-BD noise filtering process flowchart}
	\label{fig:hme}
\end{figure}

\subsection{Heterogeneous Ensemble: HTE-BD}
\label{sec:hte_bd}

\begin{algorithm}[t]
	\floatname{algorithm}{Algorithm}
	\caption{HTE-BD Algorithm}
	\label{ef}
	\begin{algorithmic}[1]
		\State \textbf{Input:} \textit{data} a RDD of tuples (label, features)
		\State \textbf{Input:} \textit{P} the number of partitions
		\State \textbf{Input:} \textit{nTrees} the number of trees for Random Forest
		\State \textbf{Input:} \textit{vote} the voting strategy (majority or consensus)
		\State \textbf{Output:} the filtered RDD without noise
		\State $partitions \gets kFold(data, P)$
		\State $filteredData \gets \emptyset$
		\ForAll{$train, test$ in $partitions$}
		\State $classifiersModel \gets learnClassifiers(train, nTrees)$
		\State $predictions \gets predict(classifiersModel, test)$
		\State $joinedData \gets join(zipWithIndex(predictions), zipWithIndex(test))$
		\State $markedData \gets$ \MAP $rf, lr, knn, orig \in joinedData$
		\State $count \gets 0$
		\IIf{$rf \neq label(orig)$} $count \gets count + 1$ \EndIIf
		\IIf{$lr \neq label(orig)$} $count \gets count + 1$ \EndIIf
		\IIf{$knn \neq label(orig)$} $count \gets count + 1$ \EndIIf
		\If {$vote = majority$}
		\IIf{$count \geq 2$} $(label = \emptyset, features(orig))$ \EndIIf
		\IIf{$count < 2$} $orig$ \EndIIf
		\Else
		\IIf{$count = 3$} $(label = \emptyset, features(orig))$ \EndIIf
		\IIf{$count \neq 2$} $orig$ \EndIIf
		\EndIf
		\ENDMAP
		\State $filteredData \gets union(filteredData, markedData)$
		\EndFor
		\State $return (filter(filteredData, label \neq \emptyset))$
	\end{algorithmic}
\end{algorithm}

Heterogeneous Ensemble is inspired by Ensemble Filter (EF)~\cite{Brodley99identifyingmislabeled}. This noise filter uses a set of three learning algorithms for identifying mislabeled instances in a dataset: a univariate decision tree (C4.5), KNN and a linear machine. It performs a $k$-fold cross validation over the training data. For each one of the $k$ parts, three algorithms are trained on the other $k-1$ parts. Each of the classifiers is used to tag each of the $test$ examples as noisy or clean.
At the end of the $k$-fold, each example of the input data has been tagged. Finally, using a voting strategy, a decision is made and noisy examples are removed.

HTE-BD follows the same working scheme as EF. The main difference is the choice of the three learning algorithms:
\begin{itemize}
	\item Instead of a decision tree, we use Spark's implementation of Random Forest.
	\item We use an exact implementation of KNN with the euclidean distance present in Spark's community repository, \mbox{kNN-IS}\footnote{\url{https://spark-packages.org/package/JMailloH/kNN_IS}} ~\cite{Maillo20173}
	\item The linear machine has been replaced by Spark's implementation of Logistic Regression, which is another linear classifier.
\end{itemize}

The noise filtering process in HTE-BD is shown in Algorithm \ref{ef}:

\begin{itemize}
	\item For each $train$ and $test$ partition of the $k$-fold performed to the input data, it learns three classification algorithms: Random Forest, Logistic Regression and 1NN using the $train$ as input data.
	\item Then it predicts the $test$ data using the three learned models. This creates a RDD of triplets $(rf, lr, knn)$ with the prediction of each algorithm for each instance.
	\item The predictions and the $test$ data are joined by index in order to compare the predictions and the original label.
	\item It compares the three predictions of each instance in the $test$ data with the original label using a Map function and, depending upon the voting strategy, the instance is marked as noise or clean.
	\item Once the Map function has been applied to each instance, noisy data is removed using a $filter$ function and the dataset is returned.
\end{itemize}

The following are required as input parameters: the dataset ($data$), the number of partitions ($P$), the number of trees for the Random Forest ($nTrees$) and the voting strategy ($vote$).

In Figure~\ref{fig:hte} we show a flowchart of the HTE-BD noise filtering process.

\begin{figure}
	\centering
	\hspace*{-1.5in}
	\includegraphics[scale=0.13]{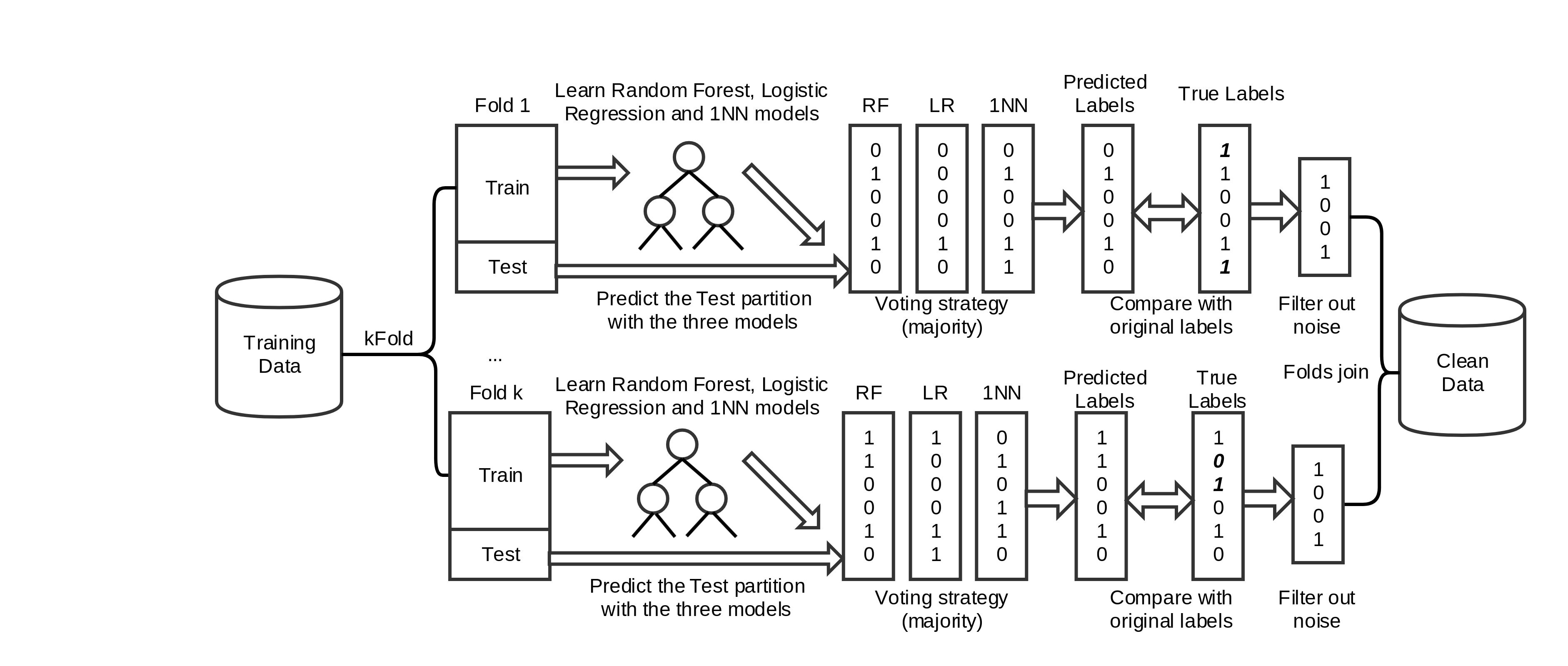}
	\caption{HTE-BD noise filtering process flowchart}
	\label{fig:hte}
\end{figure}

\subsection{Similarity: ENN-BD}
\label{sec:enn_bd}

\begin{algorithm}[t]
	\floatname{algorithm}{Algorithm}
	\caption{ENN-BD Algorithm}
	\label{ENN}
	\begin{algorithmic}[1]
		\State \textbf{Input:} \textit{data} a RDD of tuples (label, features)
		\State \textbf{Output:} the filtered RDD without noise
		\State $knnModel \gets KNN(1, "euclidean", data)$
		\State $knnPred \gets zipWithIndex(predict(knnModel,data))$
		\State $joinedData \gets join(zipWithIndex(data), knnPred)$
		\State $filteredData \gets$ \MAP $original, prediction \in joinedData$
		\If {$label(original) = label(prediction)$}
		\State $original$
		\Else
		\State $(noise, features(original))$
		\EndIf
		\ENDMAP
		\State $return (filter(filteredData, label \neq noise))$
	\end{algorithmic}
\end{algorithm}

ENN-BD is a simple filtering algorithm that works as a baseline for comparison purposes. It has been designed based on the Edited Nearest Neighbor algorithm (ENN)~\cite{wilson1972asymptotic} and follows a similarity between instances approach. ENN removes noisy instances in a dataset by comparing the label of each example with its closest neighbor. If the labels are different, the instance is considered as noisy and removed.

ENN-BD performs a 1NN using Spark's community repository kNN-IS with the euclidean distance. It checks for each instance if its closest neighbor belongs to the same class. In case the classes are different, the instance is marked as noise. Finally, marked instances are removed from the training data. This process is described in Algorithm \ref{ENN}. The only input parameter required is the dataset ($data$).

\section{Experimental Results}
\label{sec:experiments}

This section describes the experimental details and the analysis carried out to show the performance of the three noise filter methods over four huge problems.
In Section~\ref{sec:exp_framework}, we present the details of the datasets and the parameters used in the methods.
We analyze the accuracy improvements generated by the proposed framework and the study of instances removed in Section~\ref{sec:analysis}.
Finally, Section~\ref{sec:running_times} is devoted to the computing times of the proposals.

\subsection{Experimental Framework}
\label{sec:exp_framework}

Four classification datasets are used in our experiments:

\begin{itemize}
	\item SUSY dataset, which consists of 5,000,000 instances and 18 attributes. The first eight features are kinematic properties measured by the particle detectors at the Large Hadron Collider. The last ten are functions of the first eight features. The task is to distinguish between a signal process which produces supersymmetric (SUSY) particles and a background process which does not \cite{Baldi:2014kfa}.

	\item HIGGS dataset, which has 11,000,000 instances and 28 attributes. This dataset is a classification problem to distinguish between a signal process which produces Higgs bosons and a background process which does not.
	
	\item Epsilon dataset, which consists of 500,000 instances with 2,000 numerical features. This dataset was artificially created for the Pascal Large Scale Learning Challenge in 2008. It was further pre-processed and included in the LibSVM dataset repository \cite{Chang:2011:LLS:1961189.1961199}.
	
	\item ECBDL14 dataset, which has 32 million instances and 631 attributes (including both numerical and categorical). This dataset was used as a reference at the ML competition of the Evolutionary Computation for Big Data and Big Learning held on July 14, 2014, under the international conference GECCO-2014. It is a binary classification problem where the class distribution is highly imbalanced: 98\% of negative instances. For this problem, we use a reduced version with 1,000,000 instances and 30\% of positive instances.
\end{itemize}

\begin{table}[t]
	\footnotesize
	\centering
	\caption{Datasets used in the analysis}
	\label{tab:datasets}
	\begin{tabular}{ccccc}
		\toprule
		Dataset & Instances & Atts. & Total & CL \\
		\midrule 
		SUSY & 5,000,000  & 18 & 90,000,000 & 2 \\
		
		HIGGS & 11,000,000  & 28 & 308,000,000 & 2 \\
		
		Epsilon & 500,000 & 2,000 & 1,000,000,000 & 2 \\
		
		ECBDL14 & 1,000,000 & 631 & 631,000,000 & 2 \\
		\bottomrule 
	\end{tabular}
\end{table}

Table~\ref{tab:datasets} provides a brief summary of these datasets, showing the number of examples (Instances), the total number of attributes (Atts.), the total number of training data (Total), and the number classes (CL).

We carried out experiments on five levels of uniform class noise~\cite{Teng99CorrectingNoisy}: for each level of noise, a percentage of the training instances are altered by replacing their actual label by another label from the available classes. 
The selected noise levels are 0\%, 5\%, 10\%, 15\% and 20\%. 
In this case, a 0\% noise level indicates that the dataset was unaltered.
We have conducted a hold-out validation due to the time limitations of the KNN algorithm.

\begin{table}[t]
	\centering
	\caption{Parameter setting for the noise filters}
	\label{tab:params}
	\resizebox{\textwidth}{!}{
		\begin{tabular}{cp{25mm}p{85mm}}
			\toprule
			Algorithm & Parameters & Classifiers \\
			\midrule 
			HME-BD & P = 4, 5  & Random Forest: featureSubsetStrategy = "auto", impurity = "gini", maxDepth = 10 and maxBins = 32 \\
			
			HTE-BD & P = 4, 5\newline Voting = majority, consensus & 1NN, Random Forest: featureSubsetStrategy = "auto", impurity = "gini", maxDepth = 10 and maxBins = 32 \\
			
			ENN-BD & K = 1  & distance = "euclidean" \\
			\bottomrule 
	\end{tabular}}
\end{table}

\begin{table}[t]
	\centering
	\footnotesize
	\caption{Parameter setting for the classifiers}
	\label{tab:clas}
		\begin{tabular}{cp{85mm}}
			\toprule
			Classifier & Parameters \\
			\midrule 
			KNN & K = 1, distance = "euclidean" \\
			
			Decision Tree & impurity = "gini", maxDepth = 20 and maxBins = 32 \\
			
			\bottomrule 
	\end{tabular}
\end{table}

In Table~\ref{tab:params} we can see the complete list of parameters used for the noise treatment algorithms.
In order to evaluate the effect of the number of partitions on the behavior of the filters, we have selected 4 and 5 training partitions for HME-BD and HTE-BD. 
For the heterogeneous filter, HTE-BD, we also use two voting strategies: consensus (same result for all classifiers) and majority (same result for at least half the classifiers). 

Two classifiers, one  MLlib classifier, a decission tree, and one algorithm present in Spark's community repository, KNN, are used to evaluate the effectiveness of the filtering carried out by the two ensemble proposals and the similarity filter.
The decision tree can adapt its depth to avoid overfitting to noisy instances, while KNN is known to be sensitive to noise when the number of selected neighbors is low.
Prediction accuracy is used to evaluate the model's performance produced by the classifiers (number of examples correctly labeled as belonging to a given class divided by the total number of elements). 
The parameters used for the classifiers can be seen in Table~\ref{tab:clas}. Default parameters are used, except for the decision tree, in which we have tuned the depth of the tree for a better detection of noisy instances.

For all experiments we have used a cluster composed of 20 computing nodes and one master node. The computing nodes hold the following characteristics: 2 processors x Intel(R) Xeon(R) CPU E5-2620, 6 cores per processor, 2.00 GHz, 2 TB HDD, 64 GB RAM. Regarding software, we have used the following configuration: Hadoop 2.6.0-cdh5.4.3 from Cloudera's open source Apache Hadoop distribution, Apache Spark and MLlib 1.6.0, 460 cores (23 cores/node), 960 RAM GB (48 GB/node).

\subsection{Analysis of accuracy performance and removed instances}
\label{sec:analysis}

\begin{table}[t]
	\centering
	\caption{KNN test accuracy. The highest accuracy value per dataset and noise level is stressed in bold}
	\resizebox{\textwidth}{!}{
		\label{tab:knn} 
		\begin{tabular}{ll|l|ll|llll|l}
			\toprule
			Dataset & Noise (\%) & Original & HME-BD  &       & HTE-BD       &           &          &           & ENN-BD   \\
			P&            &          & 4     & 5     & 4        & 4         & 5        & 5         &       \\
			Vote&            &          &       &       & Majority & Consensus & Majority & Consensus &       \\
			\midrule
			SUSY    & 0          & 71.79    & \textbf{78.73} & 78.72 & 77.86    & 74.64     & 77.88    & 74.65     & 72.02 \\
			& 5          & 69.62    & 78.68 & \textbf{78.69} & 77.68    & 73.38     & 77.68    & 73.39     & 69.84 \\
			& 10         & 67.44    & \textbf{78.63} & 78.62 & 77.44    & 72.01     & 77.46    & 72.00     & 67.66 \\
			& 15         & 65.27    & \textbf{78.62} & 78.61 & 77.19    & 70.52     & 77.20    & 70.53     & 65.28 \\
			& 20         & 63.10    & 78.56 & \textbf{78.58} & 76.93    & 69.10     & 76.93    & 69.04     & 63.25 \\
			\midrule
			HIGGS   & 0          & 61.21    & \textbf{64.26} & 64.25 & 63.94    & 62.30     & 63.93    & 62.23     & 60.65 \\
			& 5          & 60.10    & 64.06 & \textbf{64.07} & 63.63    & 61.45     & 63.62    & 61.44     & 59.60 \\
			& 10         & 58.97    & 63.83 & \textbf{63.84} & 63.29    & 60.65     & 63.24    & 60.66     & 58.56 \\
			& 15         & 57.84    & \textbf{63.65} & 63.64 & 62.86    & 59.81     & 62.89    & 59.81     & 57.52 \\
			& 20         & 56.69    & \textbf{63.53} & 63.40 & 62.55    & 58.89     & 62.55    & 58.85     & 56.45 \\
			\midrule
			Epsilon & 0          & 56.55    & \textbf{58.11} & 58.06 & 57.43    & 55.19     & 57.39    & 55.40     & 56.21 \\
			& 5          & 55.71    & \textbf{58.64} & 58.60 & 57.47    & 55.47     & 57.39    & 55.41     & 55.43 \\
			& 10         & 55.20    & 58.51 & \textbf{58.61} & 57.26    & 55.25     & 57.26    & 55.25     & 54.79 \\
			& 15         & 54.54    & 58.39 & \textbf{58.41} & 57.00    & 55.00     & 57.02    & 55.03     & 54.30 \\
			& 20         & 54.05    & 58.02 & \textbf{58.09} & 56.75    & 54.72     & 56.71    & 54.72     & 53.68 \\
			\midrule
			ECBDL14 & 0          & 74.83    & \textbf{76.06} & 76.03 & 75.12    & 73.54     & 75.14    & 73.46     & 73.94 \\
			& 5          & 72.36    & \textbf{75.60} & 75.59 & 74.59    & 72.89     & 74.59    & 72.84     & 72.77 \\
			& 10         & 69.86    & 75.31 & \textbf{75.32} & 74.19    & 72.50     & 74.19    & 72.47     & 71.40 \\
			& 15         & 67.39    & 75.11 & \textbf{75.12} & 73.99    & 72.11     & 74.01    & 72.06     & 69.68 \\
			& 20         & 64.90    & 74.82 & \textbf{74.83} & 73.70    & 71.89     & 73.70    & 71.90     & 67.64 \\
			\bottomrule
	\end{tabular}}
\end{table}

In this section, we present the analysis on the performance results obtained by the selected classifiers after applying the proposed framework.
We denote with \emph{Original} the application of the classifier without using any noise treatment techniques, in order to evaluate the impact of the increasing noise level in the quality of the models extracted by the classification algorithms.

Table~\ref{tab:knn} shows the test accuracy values for the four datasets and the five levels of noise using the KNN algorithm for classification.
From these results we can point out that:
\begin{itemize}
	
	\item It is important to remark that the usage of any noise treatment technique always improves the \emph{Original} accuracy value at the same noise level.
Please note that the usage of the noise treatment technique allows KNN to obtain better performance at any noise level, even at the highest ones, than \emph{Original} at 0\% level for every dataset.
Since Big Datasets tend to accumulate noise, the proposed noise framework is able to improve the behavior and performance of the KNN classifier in every case.

	\item If we attend the best noise treatment strategy for KNN, we must point out that the homogeneous filter, HME-BD, enables KNN to obtain the highest accuracy values.

	\item The different number of partitions used for HME-BD has little impact in the accuracy values, which, in this respect, makes it a robust method.

	\item The heterogeneous ensemble filter, HTE-BD, is also robust to the number of partitions chosen, but its performance is lower than HME-BD.
However, the voting scheme is crucial for HTE-BD, as the consensus strategy will result in worse accuracy for KNN, being close to 2\% less accuracy for the consensus voting strategy.

	\item The baseline noise filtering method, ENN-BD, is the worst option as KNN obtains the lowest accuracy values among the three noise treatment strategies.
For ENN-BD, the accuracy drops around 2\% for each 5\% increment in noise instances. 
However, as mentioned earlier, ENN-BD is still preferable to not dealing with the noise at all.
This is due to the noise sensitive nature of KNN.

\end{itemize}

\begin{table}[t]
	\centering
	\caption{Decision tree test accuracy. The highest accuracy value per dataset and noise level is stressed in bold}
	\resizebox{\textwidth}{!}{
		\label{tab:dt}
		\begin{tabular}{ll|l|ll|llll|l}
			\toprule
			Dataset & Noise (\%) & Original & HME-BD  &       & HTE-BD       &           &          &           & ENN-BD   \\
			P&            &          & 4     & 5     & 4        & 4         & 5        & 5         &       \\
			Vote&            &          &       &       & Majority & Consensus & Majority & Consensus &       \\
			\midrule
			SUSY     & 0          & 80.24 & 79.78 & 79.79 & 79.69    & 80.27     & 79.17    & \textbf{80.29}     & 78.56 \\
			& 5          & 79.94 & 79.99 & 79.97 & 80.07    & \textbf{80.36}     & 80.10    & 80.34     & 77.49 \\
			& 10         & 79.15 & 79.85 & 79.84 & 79.81    & 80.04     & 79.81    & \textbf{80.22}     & 77.00 \\
			& 15         & 78.21 & \textbf{79.81} & 79.80 & 79.32    & 79.47     & 79.61    & 79.48     & 75.81 \\
			& 20         & 77.09 & 79.71 & \textbf{79.73} & 79.35    & 78.95     & 79.31    & 79.41     & 74.21 \\
			\midrule
			HIGGS    & 0          & 70.17 & 71.16 & \textbf{71.17} & 69.61    & 70.41     & 69.68    & 70.33     & 68.85 \\
			& 5          & 69.61 & \textbf{71.14} & 71.11 & 69.34    & 69.98     & 69.36    & 69.92     & 68.29 \\
			& 10         & 69.22 & \textbf{71.06} & 71.04 & 68.95    & 69.56     & 68.97    & 69.58     & 67.52 \\
			& 15         & 68.65 & \textbf{71.03} & 70.99 & 68.52    & 69.04     & 68.65    & 69.06     & 66.93 \\
			& 20         & 67.82 & \textbf{71.05} & 71.02 & 68.18    & 68.38     & 68.35    & 68.39     & 66.05 \\
			\midrule
			Epsilon  & 0          & 62.39 & \textbf{66.86} & 66.19 & 65.13    & 66.07     & 65.11    & 66.02     & 61.54 \\
			& 5          & 61.10 & 66.64 & \textbf{66.83} & 65.32    & 66.09     & 65.33    & 66.09     & 60.41 \\
			& 10         & 60.09 & 66.87 & \textbf{67.00} & 65.46    & 66.11     & 65.47    & 66.10     & 59.20 \\
			& 15         & 59.02 & 66.62 & \textbf{66.85} & 65.33    & 65.99     & 65.29    & 66.00     & 58.09 \\
			& 20         & 57.73 & 66.46 & \textbf{66.79} & 65.08    & 65.69     & 64.98    & 65.65     & 56.71 \\
			\midrule
			ECBDL14  & 0          & 73.98 & 74.59 & 74.38 & 74.21    & 74.51     & 74.35    & \textbf{74.62}    & 73.66 \\
			& 5          & 72.87 & 74.64 & 74.40 & 74.16    & 74.54     & 74.25    & \textbf{74.75}     & 73.48 \\
			& 10         & 71.67 & 74.59 & 74.25 & 73.84    & 74.51     & 73.94    & \textbf{74.63}     & 72.75 \\
			& 15         & 70.28 & \textbf{74.61} & 74.22 & 73.82    & 73.91     & 73.98    & 74.10     & 71.68 \\
			& 20         & 68.66 & \textbf{74.83} & 74.18 & 73.78    & 73.82     & 73.85    & 73.86     & 70.16 \\
			\bottomrule
	\end{tabular}}
\end{table}
Table~\ref{tab:dt} gathers the test accuracy values for the three noise filter methods using a deep decision tree. 
From these results we can point out that:
\begin{itemize}

	\item Again, avoiding the treatment of noise is never the best option and using the appropriate noise filtering technique will provide a significant improvement in accuracy.
	However, since the decision tree is more robust against noise than KNN, not all the filters are better than avoiding filtering noise (\emph{Original}).
	When the filters remove too many instances, both noisy and clean, the decision tree is more affected since it is able to withstand small amounts of noise while exploiting the clean instances.
	KNN was very affected by the noisy instances left, in a higher degree than the decision tree.
	Thus, a wrong filtering strategy will penalize the performance of the decision tree. We will elaborate more on this later.

	\item In terms of the best filtering technique for the decision tree, for low levels of noise, the heterogeneous ensemble HTE-BD can perform slightly better than the homogeneous HME-BD for some datasets. Nevertheless, from a 10\% noise level onwards, HME-BD outperforms HTE-BD, making it a better approach to deal with noise for the decision tree.
	
	\item Regarding the HTE-BD voting strategy, the consensus scheme achieves better results than the majority voting strategy.
Please note that the opposite has been observed in KNN: since KNN is much more sensitive and demands cleaner class borders achieved with the majority voting, the decision tree benefits from a more accurate noise removal provided by the consensus voting.
	
	\item The baseline method, ENN-BD, is achieving around 1\% less accuracy than the rest for low levels of noise, but this difference increases to 5\% less accuracy in higher noise levels. 

\end{itemize}

\begin{table}[t]
	\centering
	\caption{Average number of instances for HME-BD, HTE-BD and ENN-BD}
	\label{tab:number_instances}
	\resizebox{\textwidth}{!}{
		\begin{tabular}{l|l|l|ll|llll|l}
			\toprule
			Dataset & Noise & Original & HME-BD    &          & HTE-BD       &           &          & &   ENN-BD         \\
			P       &       &          & 4       & 5        & 4        & 4         & 5        & 5   &      \\
			Vote    &       &          &         &         & Majority & Consensus & Majority & Consensus & \\
			\midrule
			SUSY    & 0\%   & 2,500,000  & 1,984,396 & 1,983,785 & 1,974,018  & 2,281,521   & 1,973,587  & 2,280,941 & 1,262,317  \\
			& 5\%   & 2,500,000  & 1,910,750 & 1,911,317 & 1,872,868  & 2,241,766   & 1,874,053  & 2,242,598 & 1,260,781   \\
			& 10\%  & 2,500,000  & 1,837,604 & 1,837,408 & 1,801,616  & 2,207,999   & 1,800,276  & 2,203,012 & 1,258,441   \\
			& 15\%  & 2,500,000  & 1,763,890 & 1,764,176 & 1,728,789  & 2,174,051   & 1,727,949  & 2,175,876 & 1,256,611   \\
			& 20\%  & 2,500,000  & 1,691,290 & 1,691,506 & 1,657,323  & 2,144,595   & 1,657,035  & 2,141,811 & 1,254,441   \\
			
			\midrule
			HIGGS   & 0\%   & 5,500,000  & 3,900,547 & 3,900,035 & 3,567,784  & 5,048,874   & 3,564,879  & 5,051,498 & 2,765,831   \\
			& 5\%   & 5,500,000  & 3,787,000 & 3,786,366 & 3,484,271  & 5,014,344   & 3,484,274  & 5,013,132 & 2,763,942   \\
			& 10\%  & 5,500,000  & 3,672,429 & 3,672,553 & 3,404,181  & 4,972,401   & 3,401,624  & 4,973,794 & 2,760,547   \\
			& 15\%  & 5,500,000  & 3,554,120 & 3,557,252 & 3,324,547  & 4,930,575   & 3,323,465  & 4,932,060 & 2,754,636  \\
			& 20\%  & 5,500,000  & 3,446,352 & 3,443,459 & 3,242,174  & 4,888,991   & 3,240,623  & 4,886,961 & 2,756,382   \\
			
			\midrule
			Epsilon & 0\%   & 250,000   & 164,222  & 164,292  & 194,252   & 242,757    & 194,037   & 242,730  & 125,072    \\
			& 5\%   & 250,000   & 186,707  & 186,839  & 186,890   & 239,200    & 186,957   & 239,200  & 124,983    \\
			& 10\%  & 250,000   & 180,489  & 180,517  & 180,296   & 235,425    & 180,332   & 235,456  & 125,064   \\
			& 15\%  & 250,000   & 173,027  & 173,114  & 173,226   & 231,962    & 173,274   & 231,997  & 124,980    \\
			& 20\%  & 250,000   & 166,191  & 166,247  & 166,394   & 228,153    & 166,285   & 228,394  & 124,583    \\
			
			\midrule
			ECBDL14 & 0\%   & 500,000   & 387,815  & 387,873  & 393,242   & 470,731    & 393,273   & 470,924  & 367,101    \\
			& 5\%   & 500,000   & 370,991  & 371,094  & 377,451   & 458,758    & 377,239   & 459,212  & 344,717    \\
			& 10\%  & 500,000   & 357,565  & 357,270  & 361,587   & 448,460    & 361,614   & 448,550  & 324,674    \\
			& 15\%  & 500,000   & 344,363  & 344,427  & 346,454   & 439,633    & 346,633   & 439,028  & 306,832    \\
			& 20\%  & 500,000   & 330,694  & 330,761 & 331,552   & 430,444    & 331,511   & 430,357  & 292,000    \\
			
			\bottomrule
	\end{tabular}}
\end{table}

\begin{figure}[t]
	\centering
	\subfloat[SUSY]{{\includegraphics[width=5.5cm]{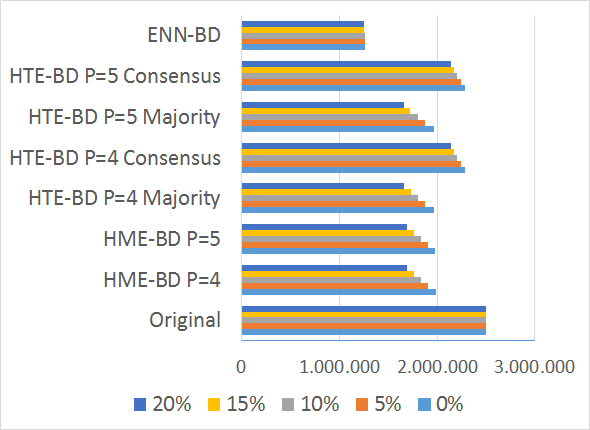} }}
	\qquad
	\subfloat[HIGGS]{{\includegraphics[width=5.5cm]{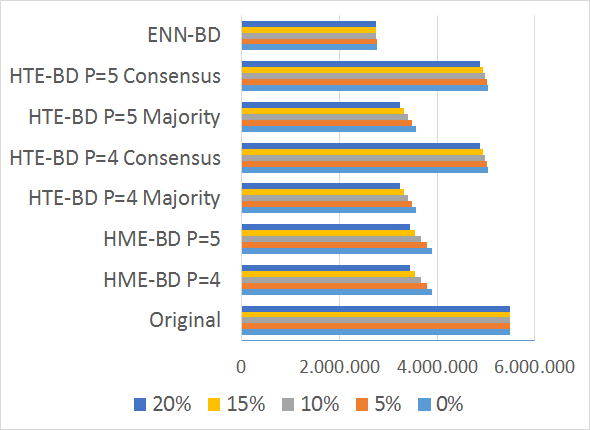} }}
	\qquad
	\subfloat[Epsilon]{{\includegraphics[width=5.5cm]{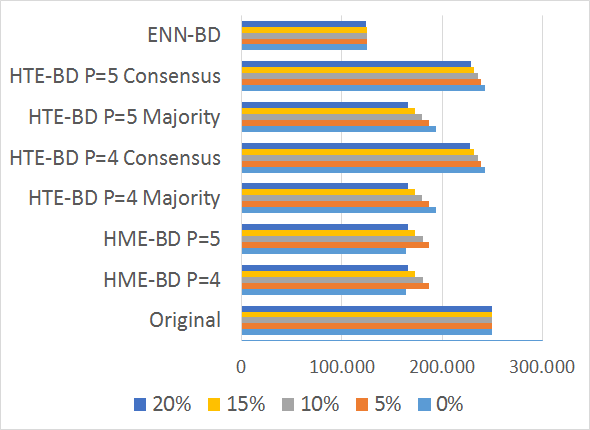} }}
	\qquad
	\subfloat[ECBDL14]{{\includegraphics[width=5.5cm]{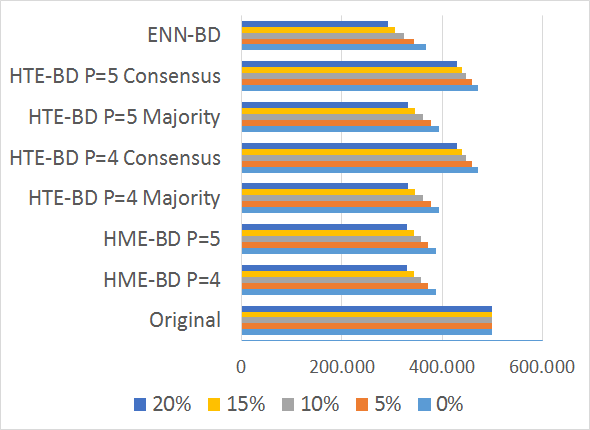} }}
	\caption{Number of instances after the filtering process}
	\label{fig:noise}
\end{figure}

The results presented have shown the importance of applying a noise treatment strategy, no matter how much noise is present in the dataset, and the best strategy overall: HME-BD.
To better explain why HME-BD is the best filtering strategy in the framework, we must study the amount of instances removed.

In Table~\ref{tab:number_instances} we present the average number of instances left after the application of the three noise filtering methods for the four datasets. 
In Figure~\ref{fig:noise} we can see a graphic representation of the number of instances for the sake of a better depiction.
As we can expect, the higher the percentage of noise, the lower the number of instances that remain in the dataset after applying the filtering technique.
However, there are different patterns depending on the filtering technique used:
\begin{itemize}
	\item For the homogeneous ensemble HME-BD, there is no effect in the number of partitions $P$ chosen with respect to the amount of removed instances. 
	On average, HME-BD removes around 20\% of the instances at a 0\% noise level. At each noise level increment an average of 3\% of the instances are removed.
	
	\item For the Epsilon dataset, at 20\% nosie, HME-BD does not remove as many instances as expected, but it is still the best option out of the two classifiers.
	A high instance redundancy in this dataset may cause homogeneous voting to not discard as many instances as the other filters.
	
	\item Like HME-BD, HTE-BD is not affected by the number of partitions, but the voting scheme does have a great impact on its behavior.
	While the majority voting strategy achieves almost the same number of removed instances as HME-BD, the consensus voting strategy is more conservative. 
	Consensus voting removes 10\% of the instances for 0\% level of noise, and it is increasing a 3\% on average as the level of noise increases, the same rate as HME-BD.
		
	\item ENN-BD is the filter that removes more instances. 
	On average it removes half the instances of the datasets for 0\% level of noise and then increases around 1\% at each increment of noise level.
	This aggresive filtering hinders the performance of noise tolerant classifiers, such as the decision tree.
	
	\item In general, HME-BD is the most balanced technique in terms of instances removed and kept.
	Although the amount of instances removed by HTE-BD with majority voting is very similar to HME-BD, the instances selected to be eliminated are different, severely affecting the classifier used afterwards. 
	
\end{itemize}

\begin{table}[t]
	\centering
	\caption{Average percentage of correctly removed instances for HME-BD, HTE-BD and ENN-BD}
	\label{tab:noise-removed}
	\resizebox{\textwidth}{!}{
		\begin{tabular}{l|l|ll|llll|l}
			\toprule
			Dataset & Noise & HME-BD &        & HTE-BD   &           &          &           & ENN-BD \\
			P       &       & 4      & 5      & 4        & 4         & 5        & 5         &        \\
			Vote    &       &        &        & Majority & Consensus & Majority & Consensus &        \\
			\midrule
			
			SUSY    & 5\%   & 79.45 & 79.44 & 78.98   & 38.16    & 79.02   & 38.18    & 50.36 \\
			& 10\%  & 76.79 & 76.74 & 77.50   & 38.03    & 77.52   & 39.24    & 50.32 \\
			& 15\%  & 76.77 & 76.77 & 76.48   & 37.71    & 76.49   & 37.71    & 50.38 \\
			& 20\%  & 79.34 & 79.37 & 78.83   & 37.26    & 78.83   & 37.25    & 50.29 \\
			\midrule
			
			HIGGS   & 5\%   & 69.19 & 69.04 & 66.04   & 26.18    & 66.03   & 26.17    & 49.12 \\
			& 10\%  & 69.26 & 69.24 & 66.10   & 25.82    & 66.09   & 25.82    & 49.56 \\
			& 15\%  & 69.39 & 69.37 & 66.14   & 25.81    & 66.14   & 25.80    & 49.62 \\
			& 20\%  & 69.45 & 69.49 & 66.23   & 25.78    & 66.23   & 25.78    & 49.55 \\
			\midrule
			
			Epsilon & 5\%   & 65.18 & 65.05 & 77.18   & 30.67    & 77.08   & 30.64    & 50.13 \\
			& 10\%  & 66.02 & 65.98 & 77.65   & 30.31    & 77.60   & 30.27    & 49.80 \\
			& 15\%  & 67.19 & 67.15 & 77.60   & 30.74    & 77.57   & 30.70    & 49.98 \\
			& 20\%  & 66.74 & 66.51 & 77.71   & 30.89    & 77.68   & 30.86    & 49.77 \\
			\midrule
			
			ECBDL14 & 5\%   & 74.45 & 74.35 & 78.30   & 44.79    & 78.28   & 45.15    & 70.83 \\
			& 10\%  & 74.36 & 74.32 & 77.26   & 46.83    & 77.22   & 46.84    & 68.55 \\
			& 15\%  & 74.41 & 74.35 & 77.07   & 44.79    & 77.04   & 44.84    & 66.45 \\
			& 20\%  & 74.38 & 74.36 & 77.29   & 43.79    & 77.26   & 43.80    & 64.25 \\
			
			\bottomrule
	\end{tabular}}
\end{table}

We have performed a deeper analysis of the removed instances, analyzing the amount of correctly removed instances for each method in the framework.

In Table~\ref{tab:noise-removed} we present the average percentage of correctly removed instances after the application of the three noise filtering methods for the four datasets. 
	In Figure~\ref{fig:noise-pct} we can see a graphic representation of these percentages of correctly removed instances.
	As we can see, the consensus voting strategy is much more conservative removing noisy instances than the rest of the methods.
	We can also outline some patterns depending on the filtering method used:

\begin{itemize}
	\item While ENN-BD is the filter that more instances removes, it is also the one that less noise removes from the datasets, averaging a 50\% of noisy instances removed.
	\item Similarly to the number of instances removed, HME-BD and HTE-BD are not affected by the number of partitions, while the voting strategy does influence the percentage of correctly removed instances.
	As we could expect, the consensus voting strategy is the one that less noisy instances clean.
	Consensus voting removes only 25\% of noisy instances in HIGGS dataset, and only increases to 45\% in ECBDL14 dataset.
	\item HME-BD and HTE-BD with majority voting, are removing aroung 65\% and 80\% of noisy instances. Both methods outperform the other in two out of four datasets.
	\item In Epsilon dataset, HTE-BD is cleaning 10\% more noisy instances than HME-BD, but HME-BD performs better in test accuracy. This can be explained by the accumulated noise of this particular dataset.
\end{itemize}

\begin{figure}[t]
	\centering
	\subfloat[SUSY]{{\includegraphics[width=5.5cm]{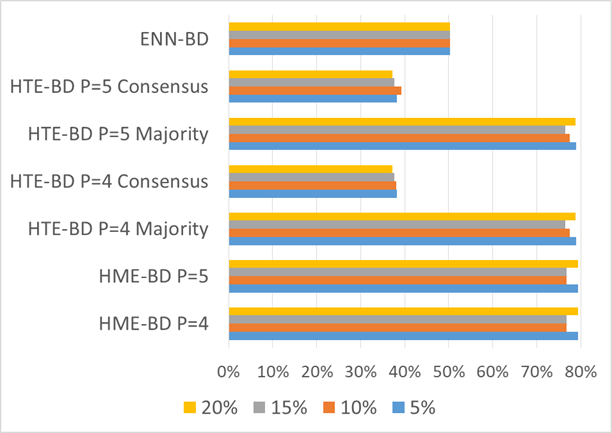} }}
	\qquad
	\subfloat[HIGGS]{{\includegraphics[width=5.5cm]{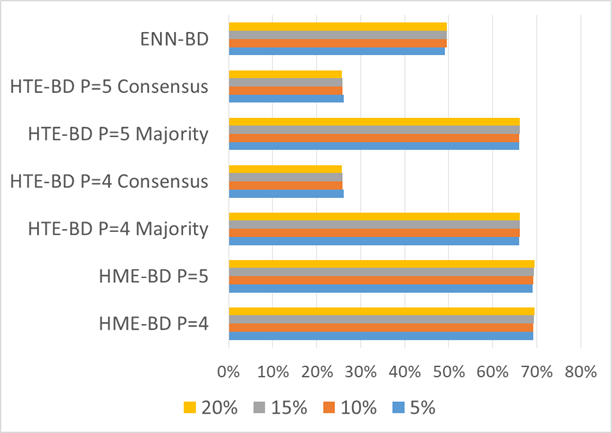} }}
	\qquad
	\subfloat[Epsilon]{{\includegraphics[width=5.5cm]{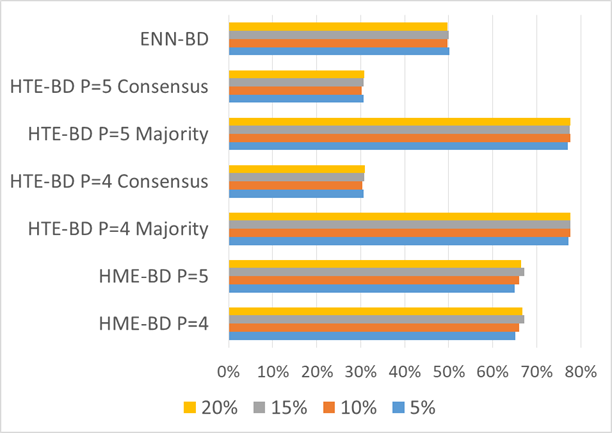} }}
	\qquad
	\subfloat[ECBDL14]{{\includegraphics[width=5.5cm]{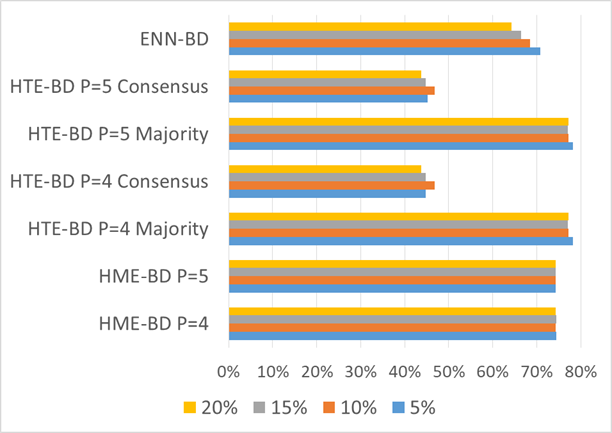} }}
	\caption{Percentage of correctly removed noisy instances after the filtering process}
	\label{fig:noise-pct}
\end{figure}

In view of the results, we can conclude that HME-BD is the most suitable ensemble option in the proposed framework to deal with noise in Big Data problems.
Even when we did not introduce any additional noise, the usage of noise treatment methods has proven to be very beneficial.
As previously mentioned, Big Data problems tend to accumulate noise and the proposed noise framework is a suitable tool to clean and proceed from Big to Smart Datasets.

\subsection{Computing times}
\label{sec:running_times}

In the previous section we have shown the suitability of the proposed framework in terms of accuracy.
In order to constitute a valid proposal in Big Data, this framework has to be scalable as well.
This section is devoted to present the computing times for the two prosposed ensemble techniques, HME-BD and HTE-BD, and the simple similarity method, ENN-BD, used as a baseline.

\begin{table}[t]
	\centering
	\caption{Average run times for HME-BD, HTE-BD and ENN-BD in seconds}
	\label{tab:times}
	\resizebox{\textwidth}{!}{
		\begin{tabular}{l|ll|llll|l}
			\toprule
			Dataset	&	HME-BD	&		&	HTE-BD	&		&		&	&	ENN-BD	\\
			P		&	4		&	5	&	4		& 	4	&	5	&	5	&\\
			Vote	&			&		&	Majority&Consensus&Majority&Consensus & \\
			\midrule 
			SUSY	&	513.46	&	632.54	&	5,511.15	&	5,855.66&6,701.62	&6,399.32&	8,956.71\\
			
			HIGGS & 587.72  & 675.07 & 15,300.62&	15,232.99 &	16,417.26 &	17,067.97 & 25,441.09\\
			
			Epsilon & 1,868.75& 2,021.14  &	4,120.79 &	7,201.05 &	5,179.09 &	5,664.06 & 2,718.97\\
			
			ECBDL14 & 1,228.24 & 1,348.10 &	9,710.70 &	11,217.02 &	10,798.18 &	11,366.01  & 14,080.03\\
			\bottomrule 
		\end{tabular}
	}
\end{table}

\begin{figure}[t]
	\centering
	\includegraphics[width=0.8\textwidth]{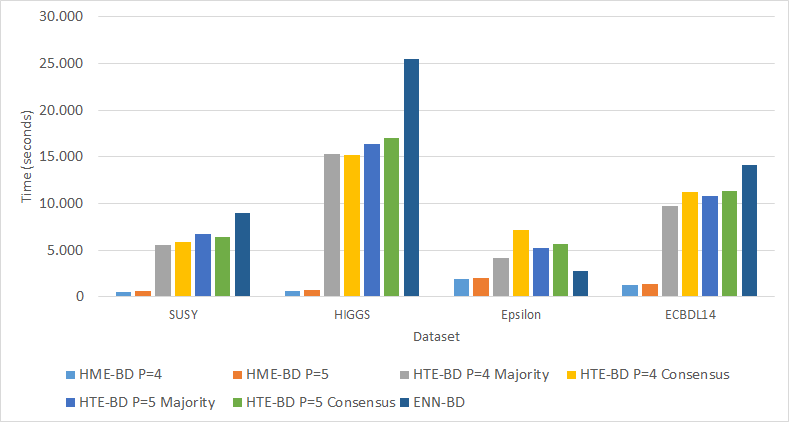}
	\caption{Run times chart}
	\label{fig:times}
\end{figure}

In Table~\ref{tab:times} we can see the average run times of the three methods for the four datasets in seconds. 
As the level of noise is not a factor that affects the run time, we show the average of the five executions performed for each dataset. 
In Figure~\ref{fig:times} we can see a graphic representation of these times.

The measured times show that the homogeneous ensemble, HME-BD, is not only the best performing option in terms of accuracy, but also the most efficient one in terms of computing time.
HME-BD is about ten times faster than the heterogeneous filter HTE-BD and the similarity filter ENN-BD.
This is caused by the usage of the KNN classifier by HTE-BD and ENN-BD, which is very demanding in computing terms.
As a result, HME-BD does not need to compute any distance measures, saving computing time and being the most recommended option to deal with noise in Big Data problems.

\section{Conclusions}
\label{sec:conclusions}

In this paper, we have tackled the problem of noise in Big Data classification, which is a crucial step in transforming such raw data into Smart Data.
We have proposed several noise filtering algorithms, implemented in a Big Data framework: Spark.
These filtering techniques are based on the creation of ensembles of classifiers that are executed in the different maps, enabling the practitioner to deal with huge datasets.
Different strategies of data partitioning and ensemble classifier combination have led to three different approaches.

The suitability of these proposed techniques has been analyzed using several data sets, in order to study the accuracy improvement, running times and data reduction rates.
The homogeneous ensemble has shown to be the most suitable approach in most cases, both in accuracy improvement and better running times.
It also shows the best balance between removing and keeping sufficient instances, being among the most balanced filter in terms of preprocessed training sets.

This work presents the first suitable noise filter in Big Data domains, where the high redundancy of the instances and high dimensional problems pose new challenges to classic noise preprocessing algorithms.
Thus, the presented framework is a valuable tool for achieving the goal of Smart Data.
It also opens promising research lines in this topic, where the presence of iterative algorithms and the usage of noise measures are also known as viable alternatives for dealing with noise.

\section*{Acknowledgment}

This work is supported by the Spanish National Research Project TIN2014-57251-P and the Foundation BBVA project 75/2016 BigDaPTOOLS.

\section*{References}

\bibliography{mybibfile}

\end{document}